\documentclass[12pt]{iopart}
\usepackage[usenames,dvipsnames]{color}
\usepackage{graphicx}
\usepackage{iopams}

\begin{document}
\title[Non-monotonic temperature dependence of chaos-assisted diffusion]{Non-monotonic temperature dependence of chaos-assisted diffusion in driven periodic systems}
\author{J. Spiechowicz$^{1,2}$, P. Talkner$^{1,3}$, P. H\"anggi$^{3,4}$ and J. {\L}uczka$^{1,2}$}
\address{$^1$ Institute of Physics, University of Silesia, 40-007 Katowice, Poland}
\address{$^2$ Silesian Center for Education and Interdisciplinary Research, University of Silesia, 41-500 Chorz{\'o}w, Poland}
\address{$^3$ Institute of Physics, University of Augsburg, 86135 Augsburg, Germany}
\address{$^4$ Nanosystems Initiative Munich, Schellingstr, 4, D-80799 M{\"u}nchen, Germany}
\ead{jerzy.luczka@us.edu.pl}
\begin{abstract}
The spreading of a cloud of independent Brownian particles typically proceeds more effectively at higher temperatures, as it derives from the commonly known Sutherland-Einstein relation for systems in thermal equilibrium. Here, we report  on a non-equilibrium situation in which the \emph{diffusion} of a periodically driven Brownian particle moving in a periodic potential \emph{decreases} with increasing \emph{temperature} within a finite temperature window. We identify as the cause for this non-intuitive behaviour a dominant deterministic mechanism consisting of a few unstable periodic orbits embedded into a \emph{chaotic} attractor together with   thermal  noise-induced dynamical changes upon varying temperature. The presented analysis is based on extensive numerical simulations of the corresponding Langevin equation describing the studied setup as well as on a simplified stochastic model formulated in terms of a three-state Markovian process. Because chaos exists in many natural as well as in artificial systems representing abundant areas of contemporary knowledge, the described mechanism may potentially be discovered in plentiful different contexts.
\end{abstract}
\maketitle
\section{Introduction}
The general idea of controlling transport is associated with the concept of implementing actions to ensure that a system exhibits desired characteristics. For example, Nature is well known in presenting sophisticated mechanisms that regulate phenomena which take place in all scales of time and space \cite{haw2007}. A careful and systematic exploration of these mechanism reveals that sometimes it is more reliable to control the setup by letting it to fluctuate and change its dynamics as little as possible to drive it to the desired state without applying intense external forces. A typical scenario Nature uses is the resource to tailored non-equilibrium forces, as it is the case for molecular and Brownian motors \cite{astumian2002,hanggi2009,spiechowicz2013jstatmech,spiechowicz2014pre, spiechowicz2016jstatmech}. Brownian machinery presents an archetypal scheme in which, even in the absence of an externally applied bias, directed motion emerges from {\it unbiased} environmental non-equilibrium perturbations via the mechanism of breaking the spatial symmetry of the setup \cite{hanggi2009}. The origin of directed transport stems from the fact that the combined action of non-equilibrium perturbations of stochastic or deterministic nature takes the system out of thermal equilibrium and breaks the detailed balance symmetry. This working principle can be seen as a key for understanding various processes occurring in numerous disciplines of science, including e.g. intracellular transport \cite{bressloff2013}. In view of the widespread applications of Brownian motors, directed motion controllability has become a focal point of research in non-equilibrium statistical physics which inspired a plethora of new microscale devices displaying unusual transport features \cite{vlassiouk2007, serreli2007, mahmud2009, costache2010, drexler2013, spiechowicz2014prb, spiechowicz2015njp, spiechowicz2015chaos, grossert2016}.

Irregular behaviour generally implies also unpredictability, which, when it comes to applications one typically attempts to avoid or at least to minimize. 
A particular aspect of unpredictability is the appearance of diffusive spreading constituting yet another salient transport quantifier. Beyond doubt, the problem of diffusion has played a central role in the development of both the foundations of thermodynamics and statistical physics \cite{hanggi2005} and has led to numerous applications within a diversity of scientific disciplines. Inspired by the physics of Brownian motors, we ask to what extent it is possible to influence the diffusive behaviour of the system in presence of thermal noise and deterministic external perturbations and elucidate the mechanisms behind such characteristics. The conventional diffusive spreading characterized by the diffusion coefficient $D$ increases with the temperature of the medium surrounding the system \cite{sutherland1905,einstein1905,smoluchowski1906}. Here, we find that the diffusion coefficient $D$ assumes a non-monotonic behaviour as a function of temperature, possessing a bell-shaped maximum at some intermediate temperature. This form of peculiar diffusive behaviour is markedly distinct from the known case of anomalous diffusion for which the mean square displacement of the particle $\langle \Delta x^2(t) \rangle$ grows asymptotically according to a power law $\langle \Delta x^2(t) \rangle \sim t^\alpha$ with $\alpha \ne 1$ \cite{metzler2014,zaburdaev2015}.  The studied non-monotonic diffusion here must also be distinguished from the phenomenon of giant diffusion. The latter has been observed in the diffusive motion of a Brownian particle in a tilted periodic landscape \cite{lindner2001, reimann2001, reimann2002, heinsalu2004, lindner2016}, in disordered potentials \cite{dan2002, garcia2014} or for an adiabatically driven inertial Brownian particle moving in a periodic one-dimensional geometry \cite{marchenko2012}.

To investigate our objective of a non-monotonic dependence of diffusion on temperature we consider the non-equilibrium dynamics of a massive Brownian particle moving in a periodic potential under the influence of a time-periodic force. The class of models described by such a dynamics includes periodically driven pendulums \cite{gitterman2010}, super-ionic conductors \cite{fulde1975}, Josephson junctions \cite{kautz1996}, dipoles rotating in external fields \cite{coffey2012}, phase-locked loops \cite{viterbi1966}, dislocations in solid state physics \cite{seeger1980}, solitons described by the Sine-Gordon equation \cite{lamb1980}, the Frenkel-Kontorova lattices \cite{braun1998}, dynamics of adatoms subjected to a time-periodic force \cite{guantes2001}, charge density waves \cite{gruner1981} and cold atoms in optical lattices \cite{denisov2014}, to name but a few.
The rich physics contained in this model has become evident over the last decades with numerous studies. In particular, with present periodic driving this class of systems comprises operational regimes that are deterministically chaotic.
The sensitive dependence on initial conditions and the abundance of unstable periodic attractors are the most salient characteristics of chaotic behaviour \cite{strogatz}.
The combination of these features, possibly assisted in addition with noise driving, make chaotic systems one of the most flexible setups enabling the emergence of the discussed peculiar diffusive behaviour.
\section{Model}
We formulate the problem in terms of a classical particle of mass $M$ which is (i) subjected to a \emph{spatially periodic} potential $U(x) = U(x + L)$ of period $L$ possessing the \emph{mirror symmetry}, i.e. there exists $x_0$ such that the relation  $U(x_0-x) =  U(x_0+x)$ holds for all $x$, (ii) additionally being agitated by an external \emph{unbiased} time-periodic deterministic force $A\cos{(\Omega t)}$ with angular frequency $\Omega$ and of amplitude strength $A$, and (iii) coupled to a thermal bath at temperature $\theta$. The overall dynamics of such a Brownian particle is determined by the following Langevin equation
\begin{equation}
	\label{eq:model}
	M\ddot{x} + \Gamma\dot{x} = -U'(x) + A\cos{(\Omega t)} + \sqrt{2\Gamma k_B \theta}\,\xi(t).
\end{equation}
Here, the dot and the prime denote differentiation with respect to the time $t$ and the particle coordinate $x$, respectively. Thermal fluctuations due to the coupling of the particle with the thermal bath are modelled by $\delta$-correlated Gaussian white noise $\xi(t)$ of zero mean and unit intensity, i.e.,
\begin{equation}
	\langle \xi(t) \rangle = 0, \quad \langle \xi(t)\xi(s) \rangle = \delta(t-s).
\label{xi}
\end{equation}
The parameter $\Gamma$ denotes the friction coefficient and $k_B$ is the Boltzmann constant. The noise intensity factor $2\Gamma k_B \theta$ with temperature $\theta$ of the heat bath follows from the fluctuation-dissipation theorem \cite{HT1982}. The latter ensures the stationarity of the thermal canonical Gibbs state for vanishing driving under the dynamics governed by Eq. (\ref{eq:model}), i.e., when $A = 0$ and in presence of periodic boundary conditions. We focus on the archetypical non-linear situation, namely, a sinusoidal potential with a barrier height $2\Delta U$,  
\begin{equation}
	U(x) = \Delta U \sin{\left( \frac{2\pi x}{L} \right)}.
\end{equation}

Upon introducing the period $L$ and the parameter combination $\tau_0 = \Gamma L^2/\Delta U$ as units for length and time, respectively, Eq. (\ref{eq:model}) is written in a dimensionless form \cite{machura2008}, reading,
\begin{equation}
	\label{eq:dimlessmodel}
	m\ddot{\hat{x}} + \dot{\hat{x}} = -\hat{U}'(\hat{x}) + a\cos{(\omega \hat{t})} + \sqrt{2Q}\,\hat{\xi}(\hat{t}),
\end{equation}
where $\hat{x} = x/L$ and $\hat{t} = t/\tau_0$. The dimensionless potential $\hat{U}(\hat{x}) = U(x)/\Delta U = U(L\hat{x})/\Delta U = \hat{U}(\hat{x} + 1)$ assumes the unit period $L = 1$ and the potential amplitude $\Delta \hat{U} = 1$. The remaining re-scaled parameters are the mass $m = (1/\Gamma\tau_0)M$, the amplitude $a = (L/\Delta U)A$ and the angular frequency $\omega = \tau_0\Omega$. The dimensionless thermal noise reads $\hat{\xi}(\hat{t}) = (L/\Delta U)\xi(t) = (L/\Delta U)\xi(\tau_0\hat{t})$ and assumes the same statistical properties, namely it is Gaussian with $\langle \hat{\xi}(\hat{t}) \rangle = 0$ and \mbox{$\langle \hat{\xi}(\hat{t})\hat{\xi}(\hat{s}) \rangle = \delta(\hat{t} - \hat{s})$}. The dimensionless noise intensity $Q = k_B\theta/\Delta U$ is given by the ratio of thermal energy $k_{B}\theta$ and half of the activation energy the particle needs to overcome the original potential barrier $2 \Delta U$. From now on, we will use only the dimensionless variables and therefore shall skip the "hat" for all quantities appearing in Eq. (\ref{eq:dimlessmodel}).

The observable of foremost interest in this study is the diffusion coefficient $D$ which characterizes the spread of trajectories and fluctuations around the average position of the particle, namely \cite{spiechowicz2015pre, spiechowicz2016scirep}, 
\begin{equation}
	\label{eq:dc}
	D = \lim_{t \to \infty} \frac{\langle [ x(t) - \langle x (t) \rangle ]^2 \rangle}{2t},
\end{equation}
where the averaging $\langle \cdot \rangle$ is over all realizations of thermal fluctuations as well as over initial conditions for the position $x(0)$ and the velocity $\dot{x}(0)$. The latter is necessary because in the deterministic limit of vanishing thermal noise intensity $Q \to 0$ the dynamics may possess several coexisting attractors thus being non-ergodic and implying that the corresponding results may be affected by a specific choice of those selected initial conditions \cite{spiechowicz2016scirep,kostur2008}.

In order to explain the discussed peculiar behaviour of the diffusion coefficient we will also investigate velocity of the Brownian particle. 
Due to the presence of the external time-periodic driving $a\cos{(\omega t)}$ as well as the friction term $\dot{x}$ in Eq. (\ref{eq:dimlessmodel}) the particle velocity $\dot{x}(t)$ approaches a unique asymptotic state in which it is characterized by a temporally periodic probability density. 
This latter density function has the same period $T$ as the driving \cite{jung1990, jung1993}. In particular, the first statistical moment of the instantaneous particle velocity $\langle \dot{x}(t) \rangle$ assumes for a long time the form
\begin{equation}
	\lim_{t \to \infty} \langle \dot{x}(t) \rangle = \langle \mathbf{v} \rangle + v(t),
\end{equation}  
where $\langle \mathbf{v} \rangle$ is the directed transport velocity. 
Note that all three terms entering the right hand side of Eq. (\ref{eq:dimlessmodel}) are \emph{unbiased}: the average of the potential force over a spatial period $L$ vanishes as well as that of the time-dependent driving over a temporal period $T$, and also the average of the random force $\xi(t)$ vanishes according to Eq. (\ref{xi}). Moreover, all three elements are symmetric and in consequence there is no directed transport in the long time stationary regime, i.e. $\langle \mathbf{v} \rangle =0$. 
The deviation $v(t)$ is periodic with period $T$ and has a vanishing time average, i.e. $v(t+T) = v(t)$ and $(1/T) \int_0^T dt\, v(t)= 0$. Therefore it is useful to consider also the {\it period averaged} velocity $\mathbf{v}(t)$ as
\begin{equation}
	\mathbf{v}(t) =  \frac{1}{T} \int_{t}^{t + T} ds\, \dot{x}(s),
\label{v}
\end{equation}
where $T = 2\pi/\omega$ denotes a period of the external driving $a\cos{(\omega t)}$. 

The particle positions $x(n T)$ at integer multiples $n$ of the period $T$ can be exactly expressed by the sequence of period averaged velocities  $\mathbf{v}(k T) = (1/T)\int_{kT}^{(k+1)T} ds\,\dot{x}(s)$ defined in Eq. (\ref{v}) as
\begin{equation}
x(n T)  = T \sum_{k=0}^{n-1} \mathbf{v}(k T)\:,
\label{xnvk}
\end{equation}
where we assumed that the particle starts at $x(0)=0$ (for $x(0)\neq 0$ one can re-define $x(nT) \to \delta x(nT) =x(nT)-x(0)$). The period averaged velocities at multiple integers of the period $T$ have a vanishing mean value. Therefore the average position remains zero and the variance of the position is given by its second moment reading
\begin{equation}
\label{xx}
\langle x^2(n T) \rangle = T^2 \sum_{k,l}^{n-1} \langle \mathbf{v}(kT)\mathbf{v}(lT) \rangle.
\end{equation}
In the asymptotic limit of large times and with the temporally periodic probability density the velocity auto-correlation function depends only on the time difference. In view of Eq. (\ref{xx}) the left hand side approaches $\langle x^2(nT)\rangle = 2nTD$ and hence  the diffusion coefficient D can be expressed  in terms of the temporal period averaged  velocity autocorrelation function, yielding
\begin{equation}
D = T \sum_{k=0}^\infty \langle \mathbf{v}(kT)  \mathbf{v}(0) \rangle.
\label{Dvv}
\end{equation}
An equivalent expression is given in \cite{MKMTHL}. The infinite time limit in Eq. (\ref{eq:dc}) implies that the auto-correlation function has to be determined for the stationary ensemble of the period averaged velocities.

Because neither the Langevin equation (\ref{eq:dimlessmodel}) nor the corresponding Fokker-Planck variant can be solved within analytical means 
we performed comprehensive numerical simulations of the model. We did so by employing a weak version of the stochastic second-order predictor-corrector algorithm with a time step typically set to about $(10^{-3} - 10^{-2}) \times T$ \cite{spiechowicz2015cpc}. Because \mbox{Eq. (\ref{eq:dimlessmodel})} is a second-order differential equation, we need to specify two initial conditions, namely $x(0)$ and $\dot{x}(0)$. 
We choose $x(0)$ and $\dot{x}(0)$ to be equally distributed over the intervals $[0, 1]$ and $[-2,2]$, respectively. Our quantities of interest were averaged over $10^3 - 10^5$ sample trajectories. All numerical calculations have been performed by use of a CUDA environment as implemented on a modern desktop GPU. This procedure did allow for a speedup of a factor of the order $10^3$ times as compared to a common present-day CPU method \cite{spiechowicz2015cpc}.
\begin{figure}[t]
	\centering
	\includegraphics[width=0.49\linewidth]{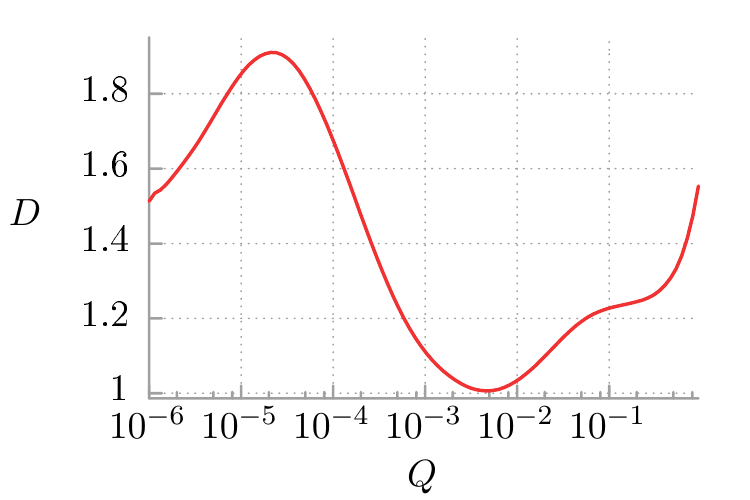}
	\caption{The dependence of the diffusion coefficient $D$  on the noise intensity 
$Q \propto \theta$ being proportional to the temperature $\theta$ of the heat-bath. The other parameters are fixed to the following values $m = 0.9, a = 8.7, \omega = 0.275$. The diffusion constant was determined from Eq. (\ref{eq:dc}) on the basis of ensembles of $10^5$ trajectories of Eq. (\ref{eq:dimlessmodel}) reaching up to $t=10^4\,T$ for different values of $Q$.}
	\label{fig1}
\end{figure}
\section{Results}
The dynamical system described by Eq. (\ref{eq:dimlessmodel}) exhibits an extremely rich behaviour as a function of the four dimensionless parameters $\{m, a, \omega, Q\}$ with dimensionless friction at $\Gamma=1$. A general overview is provided, e.g. with Ref. \cite{kautz1996}. Our objective is not the systematic exploration of the setup at hand but rather the search for the above mentioned 
{\it non-monotonic behaviour of the diffusion constant} with increasing noise intensity as shown in Fig. \ref{fig1}. The corresponding parameters of the dynamical system (\ref{eq:dimlessmodel}) are chosen as \mbox{$\{m = 0.9, a = 8.7, \omega = 0.275\}$}. At each depicted noise strength $Q \propto \theta$ the diffusion constant $D$ was estimated from $10^5$ trajectories by means of Eq. (\ref{eq:dc}). At low noise intensity $D$ increases with $Q$ until it reaches a local maximum at $Q\approx 2 \cdot 10^{-5}$. From there it decreases to a minimum at $Q\approx 5\cdot 10^{-3}$ turning over to a monotonic function of $Q$, and, finally, at sufficiently large values of $Q$ becoming strictly proportional to $Q$, i.e. to the temperature $\theta$ of the ambient thermal bath. This high temperature behaviour, however, is not depicted in Fig. \ref{fig1}. The decrease of the diffusion constant with increasing temperature $Q \propto \theta$ is counter-intuitive. It stays in clear contrast with the Einstein relation $D \propto \theta$ as well as with other known formulas, e.g. Vogel-Fulcher-like laws \cite{goychuk2014} or Arrhenius-type behaviour for the diffusion of a Brownian particle in periodic potentials \cite{lifsonjackson,festa1978,htb1990}.
\begin{figure}[t]
	\centering
	\includegraphics[width=0.49\linewidth]{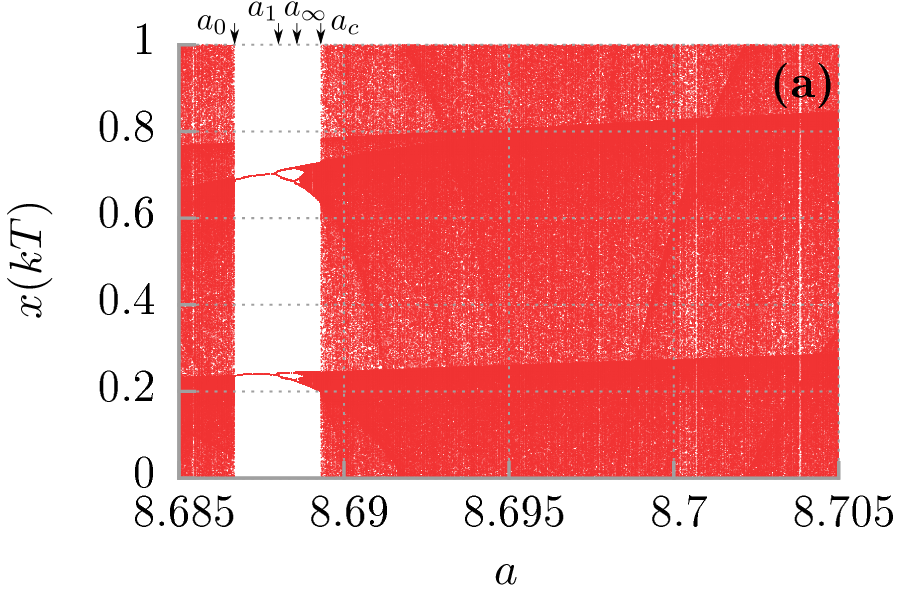}
	\includegraphics[width=0.49\linewidth]{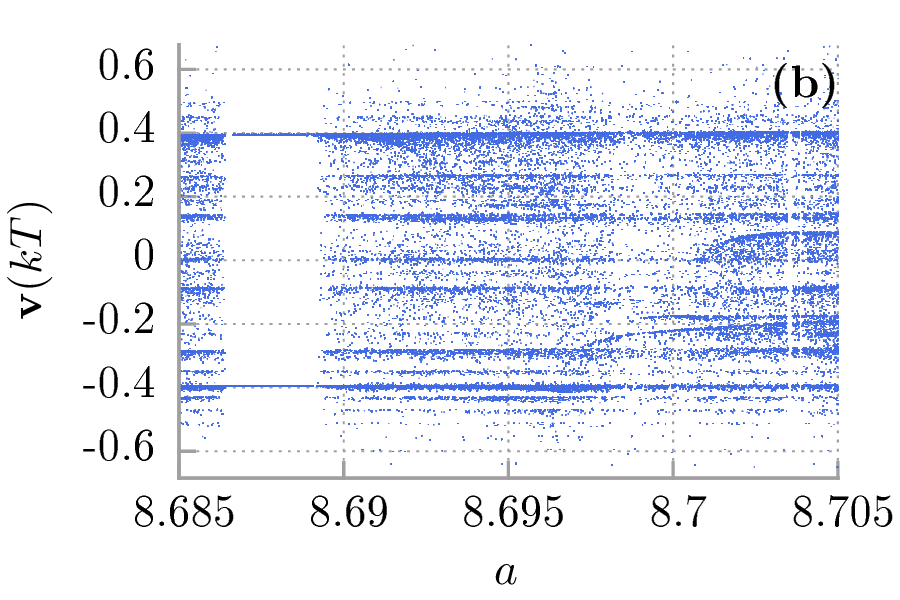}
	\caption{The deterministic dynamics in dependence of the driving amplitude $a$ is illustrated as bifurcation diagrams (a) for the stroboscopic position $x(kT)$ projected onto the principal interval $[0,L=1]$ and (b) the period averaged velocity $v(kT)$ defined in Eq. (\ref{v}) both presented after $k = 10^4$ driving cycles. 
	Chosen parameters are the same as in Fig. \ref{fig1} except now the system is deterministic $Q = 0$.
The arrows mark specific driving strengths for which the dynamics changes its character (see in text).}
	\label{fig2}
\end{figure}
\begin{figure}[t]
	\centering
	\includegraphics[width=0.49\linewidth]{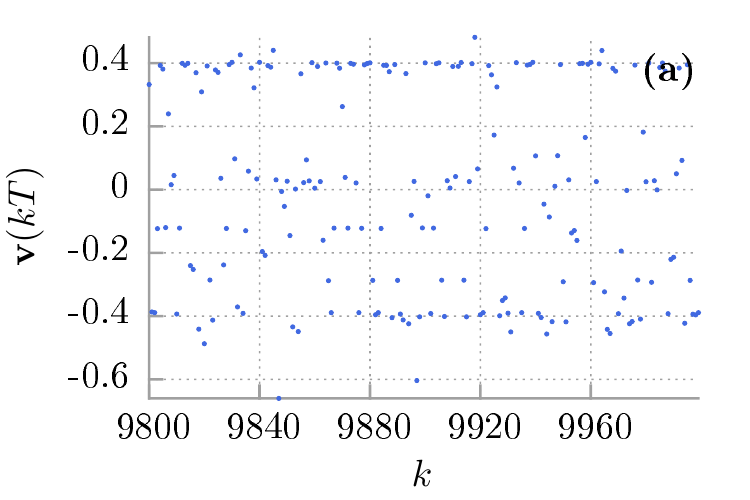}
	\includegraphics[width=0.49\linewidth]{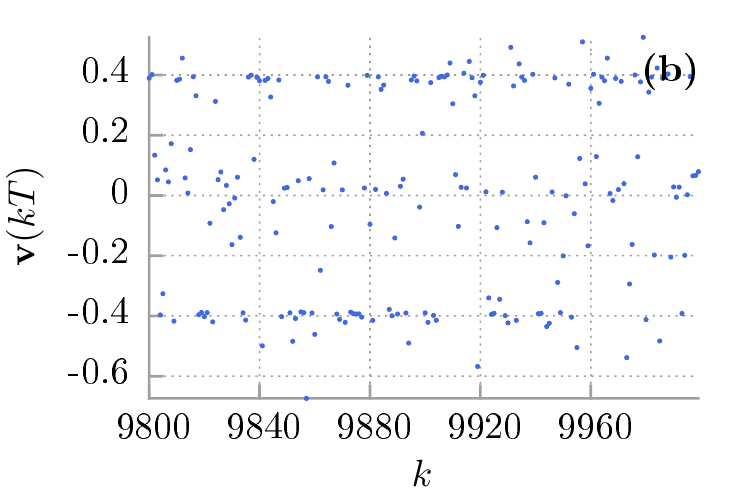}\\
	\includegraphics[width=0.49\linewidth]{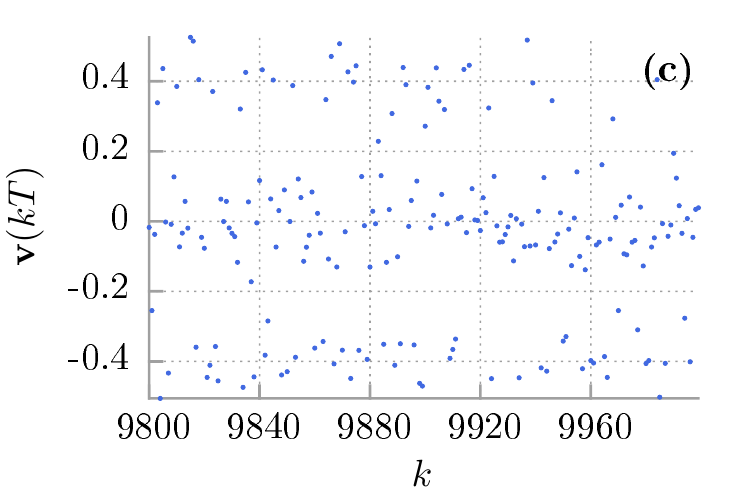}
\caption{An illustrative trajectory of the 
period averaged velocity $\mathbf{v}(kT)$, see Eq. (\ref{v}),  
is presented for the deterministic motion at $Q=0$ in panel (a) and for two different temperatures corresponding to maximal (panel (b), $Q = 2.16 \cdot 10^{-5}$) and minimal (panel (c), $Q = 0.00525$) diffusion coefficient, c.f. Fig. \ref{fig1}. Trajectories are depicted for $k=200$ driving cycles. Each blue dot represents the velocity $\mathbf{v}(kT)$ for the period $k$ at the corresponding abscissa. The remaining parameters are the same as those in Fig. \ref{fig1}. In all three cases the velocity stays predominantly close to $\mathbf{v}(kT) = \pm 0.4$ and $\mathbf{v}(kT) = 0$. In the majority of cases a trajectory does not stay for a longer period of time close to one of the dominant velocities but rather jumps in the vicinity of another one, without following any obvious rule.}
	\label{fig3}
\end{figure}
Because relevant structural elements of the stochastic dynamics described by Eq. (\ref{eq:dimlessmodel}) are determined by its deterministic properties as the first step of our analysis we consider the noiseless case.

\subsection{Noiseless dynamics: $Q=0$}

For the set of parameters presented in Fig. 1, the deterministic system (when  $Q=0$) exhibits chaotic behaviour with a dense set of unstable periodic orbits. For $Q>0$, the system 'feels' some unstable orbits which play an important role  in controlling  diffusion properties. Below we explain this. For this aim, in Fig. \ref{fig2} we present bifurcation diagrams for the positions $x(k T)$ (panel (a)) and the period averaged velocities $\mathbf{v}(kT)$ (panel (b)) as a function of the driving amplitude $a$ of the periodic driving $a\cos{(\omega t)}$ for the deterministic system (\ref{eq:dimlessmodel}) with $Q = 0$. For any given value of the driving amplitude $a$ the red dots in panel (a) represent the stroboscopic positions of the particle $x(kT)\, \mathrm{mod}\, L$, i.e. the  projected values onto the spatial principal period $[0,1]$, for $1024$ trajectories with $k=10^4$.
The velocities displayed in panel (b) show the according values $\mathbf{v}(kT)$ obtained from Eq. (\ref{v}). The bifurcation diagrams reveal three qualitatively different dynamical regimes. -- For the smallest displayed values of $a<a_0 =8.6867  $ as well as for $a>a_c=8.6893$ the motion is chaotic. The stroboscopic positions projected onto the principal spatial period almost densely cover the available interval. In both regions
the transport is diffusive \cite{inoue1982}, i.e. the diffusion constant determined by Eq. (\ref{eq:dc}) has a finite value larger than zero. In the window between $a_0$ and $a_c$ the motion is phase locked with the corresponding winding number $w = \pm 9$: the particle proceeds within one temporal period by 9 spatial periods either in positive or in negative direction depending on the initial condition. This is reflected by the two velocities $\mathbf{v}(kT) \approx \pm 0.4$ visible in panel (b) in the corresponding parameter window. In panel (a) two period doubling cascades start at the parameter value $a_1=8.6878$ and terminate at $a_\infty=8.6886$. In the small parameter interval between $a_\infty$ and $a_c$ the asymptotic motion is phase locked taking place on two coexisting chaotic attractors. The  window of phase locked motion ends at $a_c$ with an attractor merging crisis \cite{ott} giving rise to a diffusive attractor covering the full principal period. Also a wide range of velocities suddenly emerges at $a_c$ and continues to be present at larger values of the parameter $a$. However, one still detects a strong concentration of velocities at several values of $\mathbf{v}(kT)$ corresponding to integer winding numbers characterizing unstable periodic orbits. Particularly large is the probability to find  temporal period averaged velocities $\mathbf{v}(kT) \approx \pm 0.4$, corresponding to the winding numbers $w = \pm 9$. These two, together with the locked trajectories with $w = 0 $ and a few other phase locked trajectories seem to constitute the backbone of unstable periodic orbits supporting the chaotic motion. As a simplified picture of the chaotic dynamics one may think of a process in which these unstable orbits are visited in a random sequence.

\subsection{Influence of thermal noise:  $Q>0$}

In Fig.  \ref{fig3} we present the influence of thermal noise on velocity of the Brownian particle. We depict time series of period averaged velocities $\mathbf{v}(kT)$ resulting from a single trajectory in the large time limit for the same set of parameters specified in Fig. \ref{fig1}. Panel (a) exemplifies the deterministic case, with $Q=0$. Regions close to the dominant winding numbers $w = \pm 9$ and $w = 0$ are significantly more frequently visited than others. Yet, the trajectory almost never dwells near any of these states for a longer period of time but typically leaves the state within the period after it had arrived there.
Panels (b) and (c) illustrate the influence of noise on the period averaged velocity time series for the two values  $Q =2.16\cdot 10^{-5}$ and $Q=5.25 \cdot 10^{-3}$ which correspond to the positions of the maximum and the minimum of the diffusion constant $D$ displayed in \mbox{Fig. \ref{fig1}}, respectively. At the smaller noise strength displayed in panel (b) still some of the fine details of the deterministic time series are visible while others are washed out. At the larger noise intensity (panel (c)) all fine details have disappeared apart from the fact that positive and negative velocities near $\mathbf{v}(kT) \approx \pm 0.4$ still occur with relatively high probability as well as small velocities $\mathbf{v}(kT) \approx 0$.
\begin{figure}[t]
	\centering
	\includegraphics[width=0.49\linewidth]{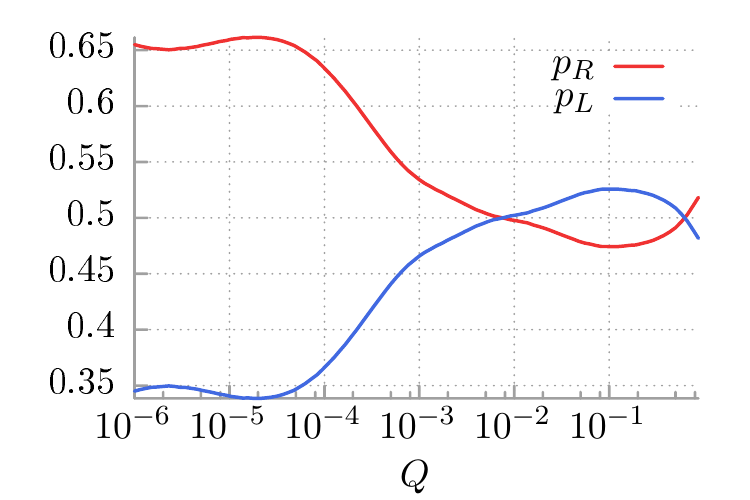}
	\caption{The probabilities $p_R$ and $p_L$ for the particle to be in the running $|\mathbf{v}(kT)| \approx 0.4$ and in the locked $\mathbf{v}(kT) \approx 0$ state, respectively, are plotted against temperature of the system $Q \propto \theta$. The diffusion coefficient $D$ is maximal when  the difference between probabilities $p_R - p_L$ has a peak, c.f. Fig. \ref{fig1}. When $p_R$ and $p_L$ intersect the diffusion coefficient $D$ is minimal. Other parameters are the same as in Fig. \ref{fig1}.}
	\label{fig4}
\end{figure}

By introducing the thresholds at $\mathbf{v}(kT) = \pm 0.2$ and counting the number of period averaged velocities which are between and outside these values one may estimate the probabilities $p_R = \mathrm{prob}(|\mathbf{v}(kT)| > 0.2)$ and $p_L = \mathrm{prob}(|\mathbf{v}(kT)| \leq 0.2)$ for the occurrence of the running and locked states, respectively. These probabilities are displayed in Fig. \ref{fig4} as a function of the noise intensity $Q$. We note that the running states occur significantly more frequent at weaker noise than at larger noise. Considering as a very rough model an independent sequence $\{v_k\}$ of velocities $\mathbf{v}(kT) \approx \pm 0.4, 0$  occurring with the respective probabilities $p_R, p_L$ one finds a diffusive behaviour for the spreading of the positions $x(n T) = \sum_{k=0}^{n-1} \mathbf{v}(k T)$ with the diffusion constant $D= 0.16 T p_R$. With the observed noise-dependence of the probability $p_R$ one already finds a qualitative agreement with the non-monotonic behaviour of the diffusion constant displayed in Fig. \ref{fig1}. However, in the next Section we will present a slightly more realistic model based on the same three velocity states but with a more adequate dynamics.
\begin{figure}[t]
\centering
\includegraphics[width=0.49\linewidth]{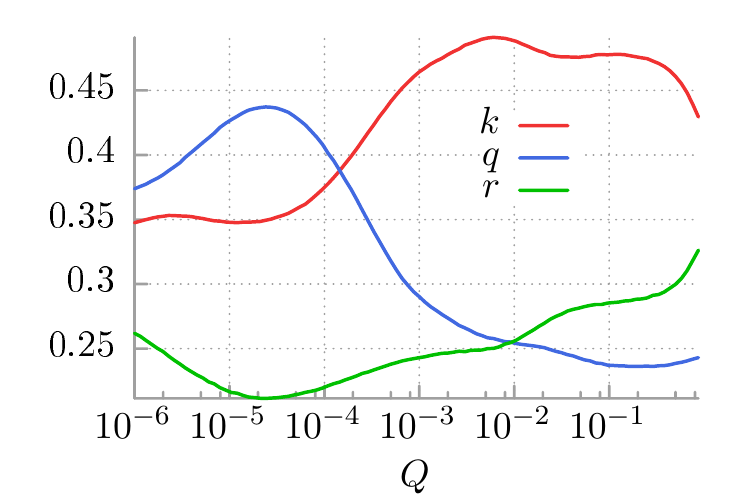}
\caption{The probabilities $k$ (red) to stay in the locked state and  $q$ (blue) to stay in either of the running states,  and the transition probability $r$ (green) to switch from a running into the locked state are displayed as functions of the noise strength $Q$. While the latter probability changes only little in the considered region of noise intensities the transition probability $k$ increases approximately in the same range of noise strengths where $q$ decreases.}
\label{fig6}
\end{figure}
\section{Approximate three-state model description of chaos induced non-monotonic diffusion}
In order to describe the non-monotonic behaviour of the diffusion constant $D$ in dependence of the noise strength $Q$ we consider a simplified model only, containing transitions between the most relevant unstable orbits. As such we consider the phase locked trajectories with winding numbers $w = \pm 9$ corresponding to the velocities $v_{\pm} \approx \pm 0.4$ and those with winding number $w = 0$ corresponding to $v_0 = 0$. The model is assumed to be symmetric with respect to the two running states $v_\pm$. The sequence of period averaged velocities $\mathbf{v}(nT)$ is then modelled by a Markov process in terms of transitions between these states. Because of the assumed symmetry the matrix $\mathbf{M}$ describing these transitions has the form
\begin{equation}
\mathbf{M}:= \left (\begin{array}{ccc}
q & \frac{1}{2}(1-k)& r\\
1-q-r &k & 1-q-r\\
r & \frac{1}{2}(1-k) & q
\end{array}
\right )\:,
\label{M}
\end{equation}
where $q$ denotes the conditional probability to remain staying in the running state $v_+$, $v_+ \rightarrow v_+$, and likewise in $v_-$, $v_- \rightarrow v_-$; further $k$, the conditional probability to remain staying in the resting state $0$, $0 \rightarrow 0$; and  $r$, the conditional probability of a transition between opposite running states $v_+ \rightarrow v_-$ and $v_- \rightarrow v_+$. Due to the assumed symmetry of the running states the transitions from the locked into a running state ($0 \rightarrow v_\pm$) is $(1-k)/2$ and from a running into the locked state ($v_\pm \rightarrow 0$) is $1-q-r$. Probabilities $q$ and $k$ may take non-negative values less than or equal to 1, while $r$ is restricted to $0\leq r \leq 1-q$. The transition probabilities can be estimated from simulations of the Langevin equation (\ref{eq:dimlessmodel}) as the relative frequencies with which the period averaged velocity $v = \mathbf{v}(nT)$ visits the three coarse grained regions $V_+=\{v| v\geq 0.2\}$,  $V_0=\{v|-0.2 < |v| < 0.2\}$ and  $V_-=\{v| v\leq -0.2\}$. The three independent rates $k$, $q$, $r$ are displayed in Fig. \ref{fig6}. Finally, in the present model the states are ``decorated'' with the respective velocities $\{v_\pm,0\}$ such that the particle changes its position within a period by $v T$ with $v\in \{v_\pm,0\}$.
\begin{figure}[t]
\centering
\includegraphics[width=0.49\linewidth]{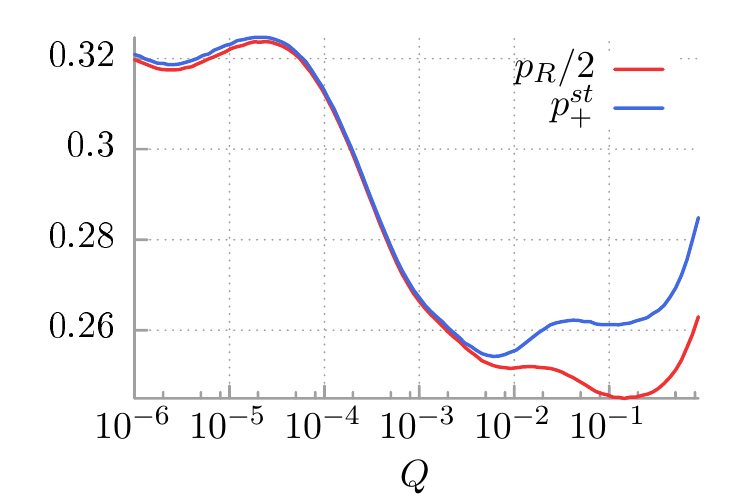}
\caption{The directly estimated probability $p_R/2$ (red) of a particle to be found in the running state as a function of the noise strength $Q$ is compared with the respective stationary probability (\ref{pst}) (blue) following from the three-state model defined by the matrix (\ref{M}) of transition probabilities. The agreement is excellent for a not too large noise intensity. For larger noise intensity the agreement is still qualitatively good.}
\label{fig7}
\end{figure}

Collecting the probabilities of finding the particle at the time $t = nT$ in either of the three states as a vector, $\mathbf{p}(n) = (p_+(n),p_0(n),p_-(n))$ one can write its dynamics as
\begin{equation}
\mathbf{p}(n+1) = \mathbf{M} p(n)
\label{ME}
\end{equation}
having the formal solution
\begin{equation}
\mathbf{p}(n) = \mathbf{M}^n \mathbf{p}(0)
\label{pn}
\end{equation}
with the initial probability $\mathbf{p}(0)$.  The stationary distribution $\mathbf{p}^{\mathrm{st}}$ is invariant under transitions, and hence is the solution of
\begin{equation}
\mathbf{p}^{\mathrm{st}} = \mathbf{M} \mathbf{p}^{\mathrm{st}}
\label{Mp}
\end{equation}
given by
\begin{equation}
\mathbf{p}^{\mathrm{st}} = \left ( \begin{array}{c}
\frac{1-k}{2(2-k-q-r)}\\
\frac{1-q-r}{2-k-q-r}\\
\frac{1-k}{2(2-k-q-r)}
\end{array}
\right )\:.
\label{pst}
\end{equation}
In Fig. \ref{fig7} the stationary probability $p^{\mathrm{st}}_+$ to find the particle in the running state  is compared with $p_R/2$ estimated from simulations of the Langevin equation (\ref{eq:dimlessmodel}).
The diffusion constant can be determined in the framework of the Markovian three-state model by means of Eq. (\ref{Dvv}) which is based on the stationary auto-correlation function of period averaged velocities. This correlation function follows from the three-state model as
\begin{equation}
\langle \mathbf{v}(n T)\mathbf{v}(0) \rangle = \sum_{\alpha,\beta} v_\alpha v_\beta
(\mathbf{M}^n)_{\alpha,\beta} p^{\mathrm{st}}_\beta\:,
\label{vv}
\end{equation}
where $\alpha$ and $\beta$ label the three velocity states and the matrix element $ (\mathbf{M}^n)_{\alpha,\beta}$ determines the conditional probability to find the state labelled by $\alpha$ after $n$ periods provided the system has started in the state $\beta$. Accordingly, $p^{\mathrm{st}}_\beta$ denotes the probability (\ref{pst}) to find $\beta$ in the stationary state. In order to evaluate the $n^{\mathrm{th}}$ power of the matrix $\mathbf{M}$ it is convenient to use its spectral representation reading
\begin{equation}
\mathbf{M} = \sum_{\kappa=1}^3 \lambda_\kappa \Pi^{(\kappa)}\:,
\label{spM}
\end{equation}
where $\lambda_\kappa$ are the eigenvalues of $\mathbf{M}$ and $\Pi^{(\kappa)}$ are matrices projecting onto the respective eigenvectors. The eigenvalue $\lambda_1 = 1$ results as a consequence of the conservation of total probability. The corresponding matrix $\Pi^{1}$  projects any vector $\mathbf{q}$ onto the stationary probability $\mathbf{p}^{\mathrm{st}}$, i.e. it acts as $\Pi^{1}\mathbf{q} = \mathbf{p}^{\mathrm{st}} \sum_\alpha q_\alpha$.
Explicit expressions of the other eigenvalues and eigenprojectors are given in the Appendix \ref{APP}. Inserting the spectral representation (\ref{spM}) into the expression (\ref{vv}) one finds that the first eigenvalue $\lambda_1$ does not contribute to the auto-correlation function because it yields a factor that is proportional to the equilibrium expectation value of the velocity and hence vanishes. Therefore the period averaged velocity auto-correlation function simplifies to read
\begin{eqnarray}
\label{vvsp}
\langle \mathbf{v}(n T)\mathbf{v}(0) \rangle &= \sum_{\kappa=2}^3 \lambda_\kappa^n \sum_{\alpha,\beta} v_\alpha v_\beta
\Pi^{(\kappa)}_{\alpha,\beta} p^{\mathrm{st}}_\beta \nonumber \\
&=2 v^2 \lambda^n_2 p^{\mathrm{st}}_1 \;,
\end{eqnarray}
where we inserted the velocities $v_1=v$, $v_2=0$ and $v_3=-v$ and made use of the explicit form of the projection matrices $\Pi^{(\kappa)}$ as given in the  \ref{APP}. Note that it does not depend on $\lambda^n_3$.  This is a consequence of symmetry of the matrix $M$ and the resulting symmetry of the eigen-projector $\Pi^{(3)}$, see equation (A6). These symmetries reflect the symmetry of the model with respect to two states with non-zero velocities $v_+$ and $v_-$. 

\begin{figure}[t]
\centering
\includegraphics[width=0.49\linewidth]{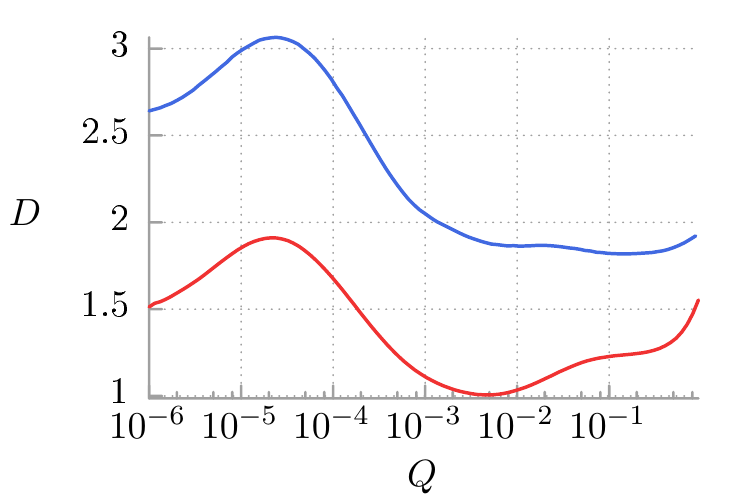}
\caption{The diffusion coefficient $D$ (red) obtained from simulation of the Langevin equation (\ref{eq:dimlessmodel}) 
is displayed as a function of the noise strength $Q$ and compared with the respective diffusion constant (blue) 
given by Eq. (\ref{D3S}) resulting from the three state model. The two curves qualitatively agree with each other in displaying the same bell-shape with matching extrema, however a quantitative discrepancy by a factor of roughly 1.5 is apparent. Possible reasons for this discrepancy are explained in the Discussion.}
\label{fig8}
\end{figure}

The sum on $n$ is readily performed to yield for the diffusion matrix
\begin{eqnarray}
D&= 2 v^2 T\frac{p^{\mathrm{st}}_1}{1-\lambda_2} \nonumber \\
&= v^2 T \frac{1-k}{(2-k-q-r)(1-q+r)} \;,
\label{D3S}
\end{eqnarray}
where we used the Eqs. (\ref{pst}) and (\ref{lambda}).
In Fig. \ref{fig8} the resulting expression is compared with the diffusion constant obtained from simulations of the Langevin equation (\ref{eq:dimlessmodel}). It turns out that (\ref{D3S}) reproduces the non-monotonic behaviour of the diffusion constant with its maximum and minimum in good agreement with the simulation result. However, the absolute magnitude of $D$ resulting from (\ref{D3S}) is too large by approximately a factor of 1.5.
\section{Discussion}
In this work we studied the spreading of a cloud of inertial, massive Brownian particles independently moving in a periodic potential. As Brownian particles they are subject to a friction as well as to a fluctuating force, both being imposed by the interaction with a thermal heat bath held at a temperature $\theta$. Moreover, they are  driven by a force periodically varying in time. In the present investigation we considered a parameter regime in which the system behaves chaotic in the absence of thermal noise, i.e., if formally the temperature is set to zero. If the temperature is gradually increased the diffusion constant characterizing the spreading rate of the particle cloud first increases until it reaches a local maximum and then decreases again until a minimum is reached. A further increase of the temperature causes a continuous growth of the diffusion constant which eventually becomes proportional to temperature.

As the reason for the counter-intuitive non-monotonic behaviour of the diffusion coefficient versus temperature we identified the dynamics of the populations of certain regions in phase space.  The latter contain certain  deterministically unstable periodic orbits. Unstable periodic orbits are known to constitute the backbone of deterministic chaos \cite{ACEGP}. They allow one to reconstruct the chaotic dynamics of a system in a hierarchical way. Using a simplified model we considered three unstable periodic orbits. Two of them move with opposite temporal period averaged velocities and a third one that is locked. The frequencies of transitions change with temperature, thereby  leading to a non-trivial dependence of the diffusion coefficient on temperature. Considering  Markovian transitions between these states with transition rates estimated from numerical simulations of the Langevin equation (\ref{eq:dimlessmodel}) then leads to a 
qualitatively agreeing description, however, with a too large diffusion coefficient in the considered range of temperatures. This discrepancy may have different reasons. First of all, non-Markovian memory effects might lead to a smaller diffusion constant. Additionally, a more systematic partitioning of the phase space \cite{badii,BFBS} and the inclusion of further unstable periodic orbits likely will  improve the agreement. Here, however, we have limited ourselves to  the generic simplest stochastic model with three states in discrete time. 
The presented analysis has been restricted to one set of the system parameters. It is not the only one and  exceptional  region where a non-monotonic dependence  of the diffusion coefficient $D$ on temperature $\theta$ is detected. We have found other regimes but are not presented here because the temperature dependence of $D$ and its  mechanisms are similar.  

Non-monotonic diffusion was also recently found in a rocked ratchet \cite{spiechowicz2015pre}; i.e. a driven Brownian particle moving in a periodic potential lacking any mirror symmetry. Due to the absence of the latter property a non-trivial directed transport effect $\langle \mathbf{v} \rangle \neq 0$ may emerge in this system. Contrary to the case considered in this paper, the deterministic limit of the mentioned regime is non-chaotic and possesses three stable coexisting states. 
The normal diffusion of the particle is caused by thermal noise produced by a heat bath in the small-to-moderate temperature limit. In the limit of weak noise a three-state modeling along the reasoning put forward here might as well provide satisfactory insight into the diffusion behavior there. 

In contrast, the present setup  exhibits chaotic motion in the deterministic limit leading already to a finite diffusion constant $D \neq 0$ at $\theta=0$. At low temperatures the combination of chaotic motion with very weak noise leads to an increase of the diffusion. With increasing noise the conditional probability to stay in the locked state eventually increases whereas that of the running state decreases going hand in hand with a decrease of the diffusion constant. For this particular mechanism to work all terms of the Langevin equation (\ref{eq:dimlessmodel}) are relevant. Without the inertial term, i.e. in the overdamped limit, the dynamics take place in a two-dimensional state space and therefore cannot display chaos. Likewise, the absence of the periodic driving also would restrict the system to a two dimensional phase space without the option of exhibiting chaos. Moreover, the periodic forcing drives the system out of thermal equilibrium in which the diffusion constant is a strictly monotonic function of temperature. For the case with vanishing friction the model becomes simple; then a dose of finite noise corresponds to infinite temperature rendering any bounded potential ineffective. In this sense with  our setup in Eq. (\ref{eq:dimlessmodel}) we deal with  a minimal model for the peculiar diffusive behaviour.

Since the latter is readily implemented in many diverse physical situations \cite{gitterman2010,fulde1975,kautz1996,coffey2012,viterbi1966,seeger1980,lamb1980,braun1998,guantes2001,gruner1981,denisov2014} we expect that a similar non-monotonic behaviour of the diffusion may be observed in various contexts recurring in the most diverse areas of science such as physics, chemistry, biology, engineering, computer science or even sociology.

We expect that our findings can be experimentally corroborated with any of the physical systems mentioned in the introductory part of the article. As a possibility we propose a promising setup for this purpose, namely the dynamics of cold atoms dwelling optical lattices \cite{lutz2013}. These systems are known for their high tunability, providing a precise control of the amplitude and period of a defect-free symmetric, spatially periodic optical potential. We consider the simplest and the most common model of a dissipative optical lattice consisting of atoms of mass $m$ with a two level structure illuminated by the optical molasses, i.e. by counter-propagating light fields with the same frequency tuned slightly below the electronic transition of the atoms. This trap generates the potential force $-U'(x)$ with precisely adjusted period and amplitude. The red-detuning of the laser fields leads to the Doppler cooling mechanism \cite{metcalf} which can be described as a classical damping term $\propto -\dot{x}$. The random photon absorption and re-emission events can be modelled by the Gaussian white noise $\xi(t)$ with noise intensity $Q$. Thus, the light beams play a role of the bath at an effective temperature to which  atoms are coupled to \cite{kindermann2016}. 
Temperature of the atoms is mainly controlled by the molasses frequency or by heating of the optical lattice due to the noisy electronic phase shifting of the beam, see Ref. 
\cite{kindermann2016} and Eq. (6) therein.

Finally, in order to generate the external driving a time dependent phase modulation $\varphi(t) = a\cos{(\omega t)}$ can be applied to one of the lattice building beams. In the laboratory reference frame such a laser configuration generates a moving optical potential. In the co-moving frame the potential is static and the atoms experience an inertial force which is proportional to the time dependent phase modulation $\varphi(t) = a\cos{(\omega t)}$ \cite{renzoni2003, wickenbrock2012}. In this way, the predicted non-monotonic behaviour of the diffusion constant as a function of temperature should manifest itself in an optical lattice experiment. Last but not least, we stress that the detected peculiar diffusive behaviour manifesting in non-monotonic temperature dependence of the diffusion coefficient can be observed in \emph{deep} optical lattices. This non-intuitive diffusion behaviour is therefore quite distinct from the well explored regime with shallow potentials where typically anomalous diffusion processes occur \cite{lutz2003, lutz2004}.
\section*{Acknowledgements}
This work was supported in part by the MNiSW program via  the Diamond Grant (J.S.) and the NCN grant 2015/19/B/ST2/02856 (J. {\L}). P.T. thanks the  Foundation for Polish Science (FNP) for granting him an Alexander von Humboldt Honorary Research Fellowship.
%
%
%
\appendix
\section{Spectral representation of $\mathbf{M}$}\label{APP}
The spectral representation (\ref{spM}) of the matrix $\mathbf{M}$ defined in Eq. (\ref{M}) is given by the eigenvalues $\lambda_\kappa$ reading
\begin{eqnarray}
\lambda_1 &=1 \\
\lambda_2 &= q-r \\
\lambda_3 &=q+k+r-1
\label{lambda}
\end{eqnarray}
and the projection matrices
\begin{eqnarray}
\Pi^{(1)} & = \frac{1}{2-q-k-r} \left ( \begin{array}{ccc}
\frac{1}{2} (1-k) &\frac{1}{2} (1-k) &\frac{1}{2} (1-k) \\
1-q-r &1-q-r &1-q-r \\
\frac{1}{2} (1-k) &\frac{1}{2} (1-k) &\frac{1}{2} (1-k)
\end{array}
\right ) \\
\ \ \ \nonumber  \\
\Pi^{(2)}&=\frac{1}{2} \left ( \begin{array}{ccc}
1 &0 & -1\\
0&0&0\\
-1 &0&1
\end{array}
\right )\\
\ \ \ \nonumber \\
\Pi^{(3)}&=\frac{1}{2 (2-q-k-r)} \left ( \begin{array}{ccc}
1-q-r& -1+k& 1-q-r\\
-2(1-q-r) & 2(1-k) & -2 (1-q-r)\\
1-q-r& -1+k& 1-q-r
\end{array}
\right ) \hspace{1.0cm}
\end{eqnarray}
\label{Pi}

\end{document}